\documentclass{PoS}
\usepackage{subcaption}

\def\slash#1{\mbox{$\not \!\! #1$}}
\def\Dslash{{\slash {\cal D}}}

\def\lvec#1{\setbox0=\hbox{$#1$}
    \setbox1=\hbox{$\scriptstyle\leftarrow$}
    #1\kern-\wd0\smash{
    \raise\ht0\hbox{$\raise1pt\hbox{$\scriptstyle\leftarrow$}$}}
    \kern-\wd1\kern\wd0}
\def\rvec#1{\setbox0=\hbox{$#1$}
    \setbox1=\hbox{$\scriptstyle\rightarrow$}
    #1\kern-\wd0\smash{
    \raise\ht0\hbox{$\raise1pt\hbox{$\scriptstyle\rightarrow$}$}}
    \kern-\wd1\kern\wd0}

\def\diracstar#1#2{
    \setbox0=\hbox{$\gamma$}\setbox1=\hbox{$\gamma_{#1}$}
    \gamma_{#1}\kern-\wd1\kern\wd0
    \smash{\raise4.5pt\hbox{$\scriptstyle#2$}}}

\def\tr{\,\hbox{tr}\,}

\newcommand{\beq}{\begin{equation}}
\newcommand{\eeq}{\end{equation}}
\newcommand{\beqn}{\begin{eqnarray}}
\newcommand{\eeqn}{\end{eqnarray}}

\makeatletter
\renewcommand\subsection{\@startsection{subsection}{2}{\z@}%
                                     {-3.25ex\@plus -1ex \@minus -.2ex}%
                                     {-1.5ex \@plus .2ex}%
                                     {\normalfont\large\bfseries}}
\renewcommand\subsubsection{\@startsection{subsubsection}{3}{\z@}%
                                     {-3.25ex\@plus -1ex \@minus -.2ex}%
                                     {-1.5ex \@plus .2ex}%
                                     {\normalfont\normalsize\bfseries}}
\makeatother
\title{Non-perturbative generation of elementary fermion
mass: a numerical study
}

\ShortTitle{Non-perturbative generation of elementary fermion
masses: a numerical study
}

\author{
S.\ Capitani$^{a)}$, G.M.\ de Divitiis$^{b)}$, P.\ Dimopoulos$^{c)}$, R.\ Frezzotti$^{b)d)}$,
 \speaker{M.\ Garofalo}$^{d)}$, 
 B.\ Kostrzewa$^{e)}$, F.~Pittler$^{e)}$ , G.C.\ Rossi$^{b)c)}$,  C.\ Urbach$^{e)}$
\\
 $^{a)}$ Johann Wolfgang Goethe-Universit\"at Frankfurt am Main
Institut f\"ur Theoretische Physik
Max-von-Laue-Stra\ss e 1
D-60438 Frankfurt am Main
Germany
\\
$^{b)}$ 
  Dipartimento di Fisica, Universit\`a di  Roma
  ``{\it Tor Vergata}'' and INFN, Sezione di Roma 2,
     Via della Ricerca Scientifica - 00133 Rome, Italy\\
 $^{c)}$ Centro Fermi - Museo Storico della Fisica e Centro Studi e Ricerche Enrico Fermi, Compendio del Viminale, Piazza del Viminiale 1, I-00184, Rome, Italy\\
  $^{f)}$  Istituto Nazionale di Fisica Nucleare (INFN), Sezione di Roma Tre, 00146 Rome, Italy
   \\
 $^{e)}$ 
 Institut f\"ur Strahlen-und Kernphysik (Theorie), Nussallee 14-16 Bethe Center for Theoretical Physics, Nussallee 12 Universit\"at Bonn, D-53115 Bonn, Germany\\
\\
        E-mail: \email{marco.garofalo@roma2.infn.it}}

\abstract{In this talk we present a numerical lattice study of a SU(3) gauge model where a SU(2) doublet of non-Abelian strongly interacting fermions is coupled to a complex scalar field doublet via a Yukawa and a Wilson-like term. The model enjoys an exact symmetry, acting on all fields, which prevents UV power divergent fermion mass corrections, despite the presence of these two chiral breaking operators in the Lagrangian. In the phase where the scalar potential is non-degenerate and fermions are massless, the bare Yukawa coupling can be set at a critical value at which chiral fermion transformations become symmetries of the theory. Numerical simulations in the Nambu-Goldstone phase of the critical theory, for which the renormalized Yukawa coupling by construction vanishes, give evidence for non-perturbative generation of a UV finite fermion mass term in the effective action.}

\FullConference{The 36th Annual International Symposium on Lattice Field Theory - LATTICE2018\\
		22-28 July, 2018\\
		Michigan State University, East Lansing, Michigan, USA.}

\begin{document}
\section{Introduction}
In~\cite{Frezzotti:2014wja} a new non-perturbative (NP) mechanism for elementary
particle mass generation was conjectured.
In this contribution we provide numerical evidence that this
mass generation mechanism  is realized
in the simplest  $d = 4$ gauge
model where this phenomenon can take place.
The Lagrangian of the model is
\beqn
&&{\cal L}_{\rm{toy}}(\Psi,A,\Phi)= {\cal L}_{kin}(\Psi,A,\Phi)+{\cal V}(\Phi)
+{\cal L}_{Wil}(\Psi,A,\Phi) + {\cal L}_{Yuk}(\Psi,\Phi) \, ,\label{SULL} \\
&&\quad{\cal L}_{kin}(\Psi,A,\Phi)= \frac{1}{4}(F\cdot F)+\bar \Psi_L\Dslash \Psi_L+\bar \Psi_R\Dslash \,\Psi_R+\frac{1}{2}{\tr}\big{[}\partial_\mu\Phi^\dagger\partial_\mu\Phi\big{]}\label{LKIN}\\
&&\quad{\cal V}(\Phi)= \frac{\mu_0^2}{2}{\tr}\big{[}\Phi^\dagger\Phi\big{]}+\frac{\lambda_0}{4}\big{(}{\tr}\big{[}\Phi^\dagger\Phi\big{]}\big{)}^2\label{LPHI}\\
&&\quad{\cal L}_{Wil}(\Psi,A,\Phi)= \frac{b^2}{2}\rho\,\big{(}\bar \Psi_L{\overleftarrow{\cal D}}_\mu\Phi {\cal D}_\mu \Psi_R+\bar \Psi_R \overleftarrow{\cal D}_\mu \Phi^\dagger {\cal D}_\mu \Psi_L\big{)}
\label{LWIL} \\
&&\quad{\cal L}_{Yuk}(\Psi,\Phi)=\
  \eta\,\big{(} \bar \Psi_L\Phi \Psi_R+\bar \Psi_R \Phi^\dagger \Psi_L\big{)}
\label{LYUK} \, ,
\eeqn
where $b^{-1}=\Lambda_{UV}$ is the UV-cutoff.
The Lagrangian~(\ref{SULL}) describes a
SU(2) fermion doublet subjected to non-Abelian gauge interaction and coupled to a complex scalar field via Wilson-like
(eq.~(\ref{LWIL})) and Yukawa (eq.~(\ref{LYUK})) terms. For short we use a compact SU(2)-like notation where 
$\Psi_L=(u_L\,\,d_L)^T$ and $\Psi_R=(u_R\,\,d_R)^T$ are fermion iso-doublets and $\Phi$ is a $2\times2$
matrix with $\Phi=(\phi,-i\tau^2 \phi^*)$ and $\phi$ an iso-doublet of complex scalar fields.
The term ${\cal V}(\Phi)$ in eq.~(\ref{LPHI}) is the standard quartic scalar potential where the 
(bare) parameters $\lambda_0$ and $\mu_0^2$ control the self-interaction and the mass of the scalar
field. In the equations above we have introduced the covariant derivatives
\beq
{\cal D}_\mu=\partial_\mu -ig_s \lambda^a A_\mu^a \, , \qquad
\overleftarrow{\cal D}_\mu =\overleftarrow{\partial}_\mu +ig_s \lambda^a A_\mu^a \, ,\label{COVG}
\eeq
where $A_\mu^a$ is the gluon field ($a=1,2,\dots, N_c^2-1$) with field strength $F_{\mu\nu}^{a}$, here $N_c=3$.
Among other symmetries, the Lagrangian~(\ref{SULL}) is
invariant under the global transformations ($\Omega_{L/R} \in {\mbox{SU}}(2)$)
\begin{eqnarray}
\hspace{-.5cm}&&\chi_L\times \chi_R =  [\tilde\chi_L\times (\Phi\to\Omega_L\Phi)]\times [\tilde\chi_R\times (\Phi\to\Phi\Omega_R^\dagger)] \, ,\label{CHIL}\\
\hspace{-.5cm}&&\tilde\chi_{L/R} : Q_{L/R}\rightarrow\Omega_{L/R} Q_{L/R} \, ,\quad \bar Q_{L/R}\rightarrow \bar Q_{L/R}\Omega_{L/R}^\dagger \, . \label{GTWT}
\end{eqnarray}
No mass like term of the form $\bar Q_L Q_R+ \bar Q_R Q_L $ can be generated as it is not $\chi_L\times\chi_R$ invariant.  \\
The Lagrangian~(\ref{SULL}) is not invariant under the
purely fermionic chiral transformations $\tilde\chi_L\times\tilde\chi_R$.
However,  in the phase with positive renormalized squared scalar mass ($\hat\mu^2_\phi>0$), where 
 the $\chi_L\times\chi_R$ symmetry is realized  {\it \`a la} Wigner, 
  a critical value $\eta_{cr}$ of the bare Yukawa parameter exists where the effective Yukawa term can be made to vanish \cite{Frezzotti:2014wja}.
This critical value can be identified looking at the renormalized Schwinger-Dyson equation (SDE) of the Axial $\tilde\chi_L\otimes \tilde \chi_R$ transformations
\begin{equation}
 Z_{\tilde A}\partial_\mu \langle \tilde J^{A,i}_\mu(x) \,\hat {\cal O}(0)\rangle = 
[\bar\eta(\eta,g_s,\rho,\lambda_0)-\eta ] \,\langle \Big{[} \overline Q_L\left\{\frac{\tau^i}{2},\Phi\right\} Q_R-\bar Q_R\left\{\frac{\tau^i}{2},\Phi^\dagger\right\} Q_L\Big](x)\,\hat{\cal  O}(0)\rangle +\mbox{O}(b^2) \,,
\label{A_SDE}
\end{equation}
where  $\tilde J_\mu^{A,i}=  \overline Q_L\gamma_\mu\gamma_5\frac{\tau^i}{2}Q_L $ is the axial current and $Z_{\tilde A}$ its renormalization constant.
$\bar \eta$  is a $\mu_0^2$ independent dimensionless function that controls the mixing of the Wilson-like term with the Yukawa operator.
Setting $\eta=\eta_{cr}(g^2_s,\rho,\lambda_0)$ such that $\eta_{cr}(g^2_s,\rho,\lambda_0)-\bar\eta(\eta_{cr}; g^2_s,\rho,\lambda_0)=0$
the SDEs take the form of 
Ward--Takahashi identities (WTI) and the fermionic chiral 
transformations $\tilde\chi_L\times\tilde\chi_R$ become 
(approximate) symmetries~\cite{Bochicchio:1985xa} of the model~(\ref{SULL}).

The situation where $\hat\mu^2_\phi<0$ and
${\cal V}(\Phi)$ has a double-well shape with 
$\langle \Phi \rangle= v$, and the 
$\chi_L\times\chi_R$ symmetry is realized {\it \`a la} Nambu-Goldstone (NG) is physically more interesting. 
In this phase a non-perturbative term is expected/conjectured  
\cite{Frezzotti:2014wja} to appear
in the effective Lagrangian of the theory. Indeed
the occurrence of the  non-perturbative term 
can be interpreted
by stipulating that 
  the effective action contains the term  
$\Gamma\supset{c_1\Lambda_s[ \overline Q_L { U} Q_R+\mbox{hc}]}$,
which provide a mass for the fermions.
Here $\Lambda_s$ is the RGI scale of the theory and $U=\Phi/\sqrt{\Phi^\dagger \Phi}$ is a  dimensionless effective field transforming as $U \rightarrow \Omega_L U \Omega_R^\dagger$ 
under $\chi_L \times \chi_R$, 
it
represents the exponential Goldstone boson
map and makes sense only if $v\neq 0$ \cite{Frezzotti:2014wja},
the latter term provide a mass for the fermions. 

\section{Numerical investigation}
\label{Numerical investigation}
The arguments given in Ref.~\cite{Frezzotti:2014wja} imply that the NP mechanism
leading to elementary fermion mass generation (if any) occurs even if
virtual fermion effects are neglected, then in this first investigation we decided to
work in the quenched fermion approximation, thus gauge and scalar configurations can be generated independently from each other.
The details of the lattice regularization of~(\ref{SULL}) we use in our numerical study can be found in~\cite{Capitani:2017trq}. 
Due to the exact $\chi_L\times\chi_R$ symmetry the lattice artefacts are only O$(b^{2n})$.
In the present quenched study exceptional configurations of the gauge
fields and the scalars  can occur in the Monte Carlo sampling leading
to small eigenvalues of lattice Dirac operator.
In order to control exceptional configurations we add a twisted mass term in the action
 \begin{equation}
 S_{lat}^{toy+tm}=S_{lat}+i\mu b^4\sum_x\overline \Psi \gamma_5\tau_3\Psi
 \end{equation}
  at the price of introducing a soft (hence harmless) breaking of $\chi_{L,R}$ (and $\widetilde \chi_{L,R}$ when restored). All the results will be extrapolated to $\mu\to0$. 

We simulate at three different values of the bare gauge coupling $\beta= 5.75,\, 5.85$ and $5.95$, which 
correspond to three different lattice spacings 0.15, 0.12 and 0.10~fm, assuming for the 
Sommer scale fixed $r_0 = 0.5$~fm \cite{Guagnelli:1998ud,Necco:2001xg}.
The coefficient of the Wilson-like parameter in eq.~(\ref{LWIL}) is kept fixed at $\rho=1.96$.
In the NG phase for the three lattice spacing  the parameters of the scalar sector are fixed by the renormalization conditions 
$M^2_\sigma r_0^2\!= \!1.284$ and   $\lambda_{R}\!\equiv\!M_\sigma^2/(2v_R^2)\!=\!0.441$ where $M_{\sigma}$ is the mass of the (non-Goldstone boson) scalar particle.
We find that the values of the bare parameters that satisfy the renormalization condition are $(bm_{\phi})^2= -0.5941$, $-0.5805$, $ -0.5756$ and $\lambda_0=0.5807$, $0.5917$, $0.6022$ respectively.
 The phase in which the theory lives is selected by taking the bare scalar mass $m_{\phi}^2$ 
  either larger (Wigner) or smaller
(NG phase) than $m_{cr}^2$, where $m_{cr}^2$ is the value of $m_\phi^2$ at
which the scalar susceptibility diverges in the infinite volume limit.

\subsection{Wigner phase:}
\begin{figure}
\hspace{-0.3cm}
\begin{subfigure}{.33\textwidth}
\includegraphics[scale=0.38, trim=90 -20 0 0]{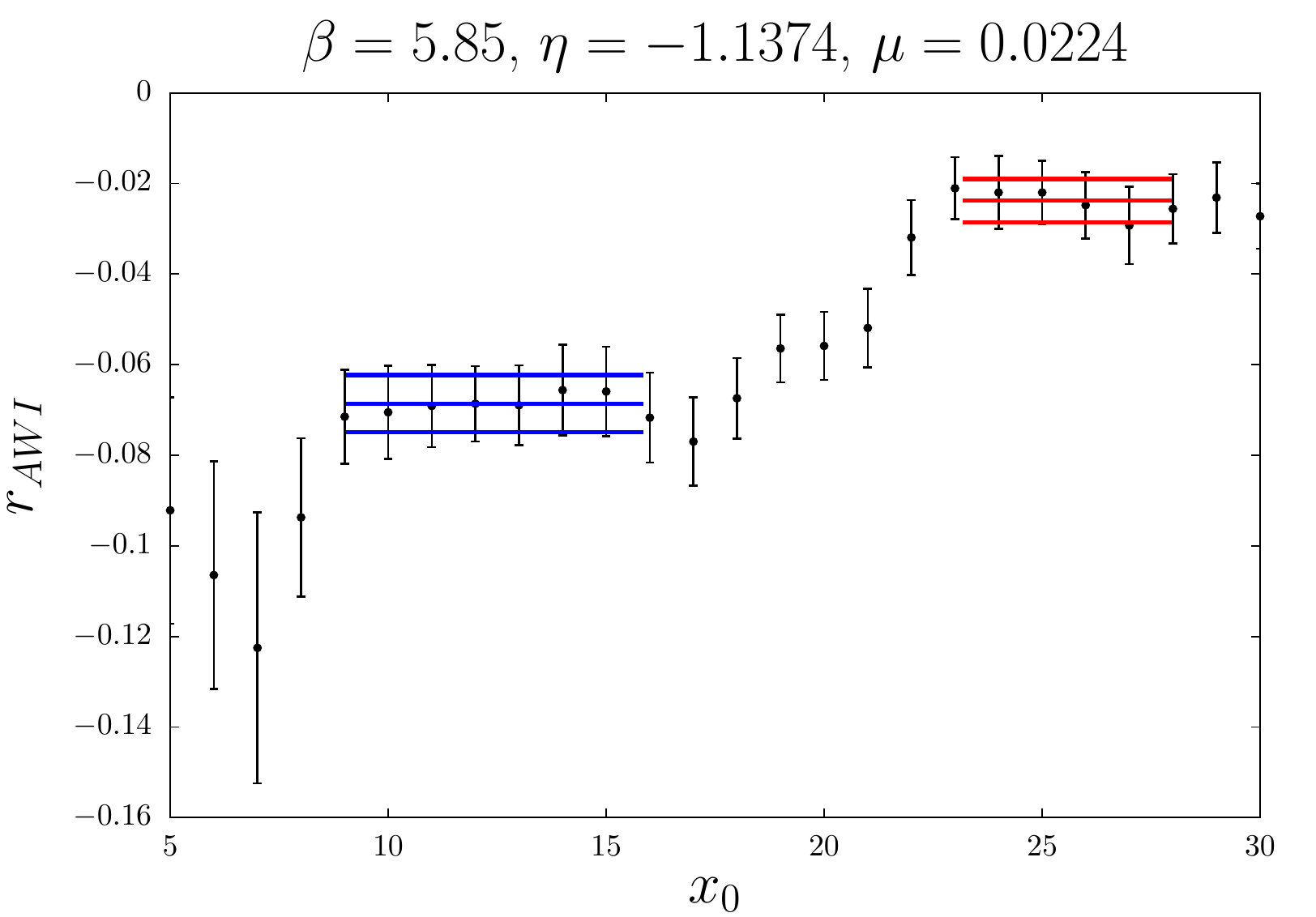}
\caption{}
\label{r_AWI_plateau}
\end{subfigure}
\begin{subfigure}{.33\textwidth}
\includegraphics[scale=0.45, trim=10 0 0 0]{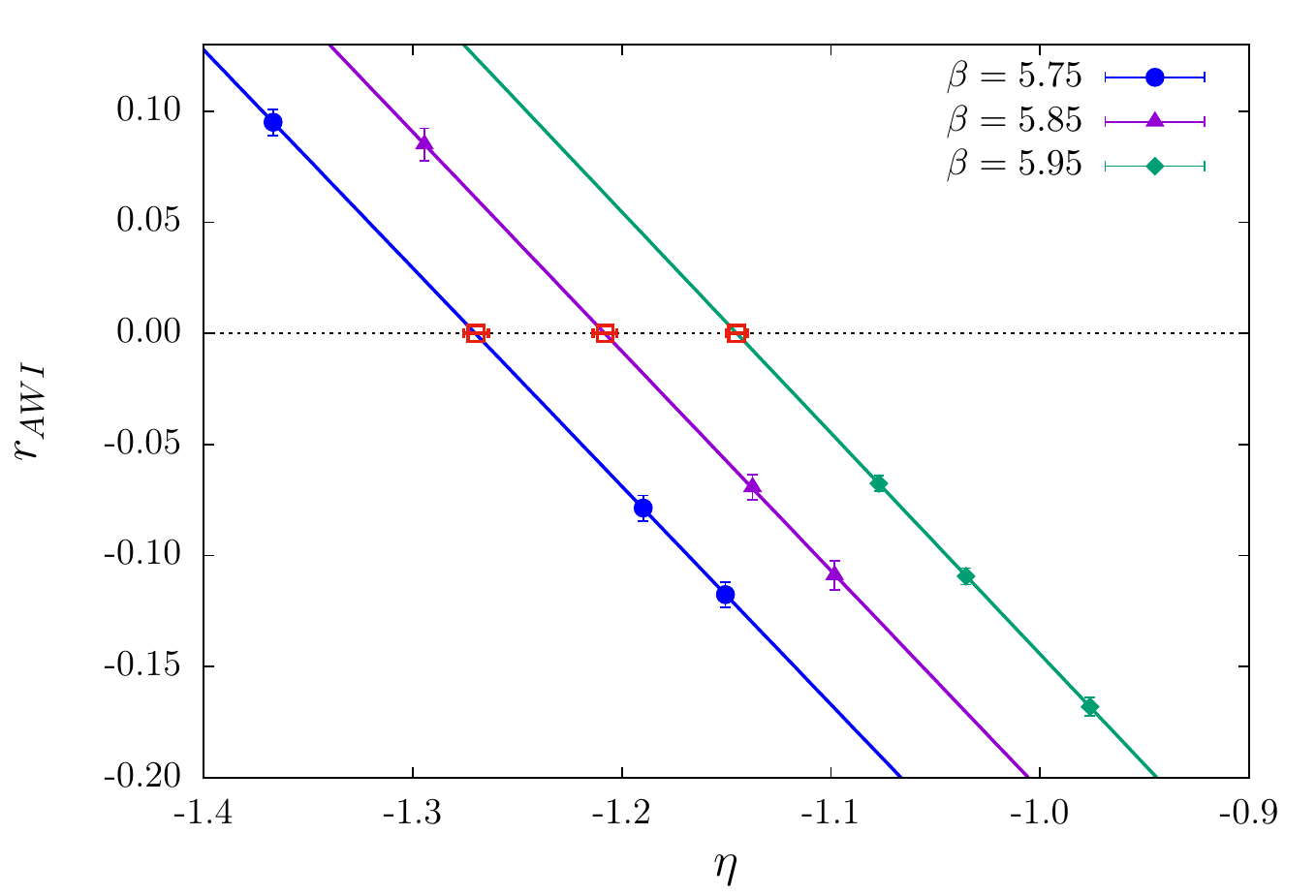}
\caption{}
\label{fig:figRAWI}
\end{subfigure}
\hspace{0.1cm}
\begin{subfigure}{.33\textwidth}
\includegraphics[scale=0.35, trim=-50 0 0 0]{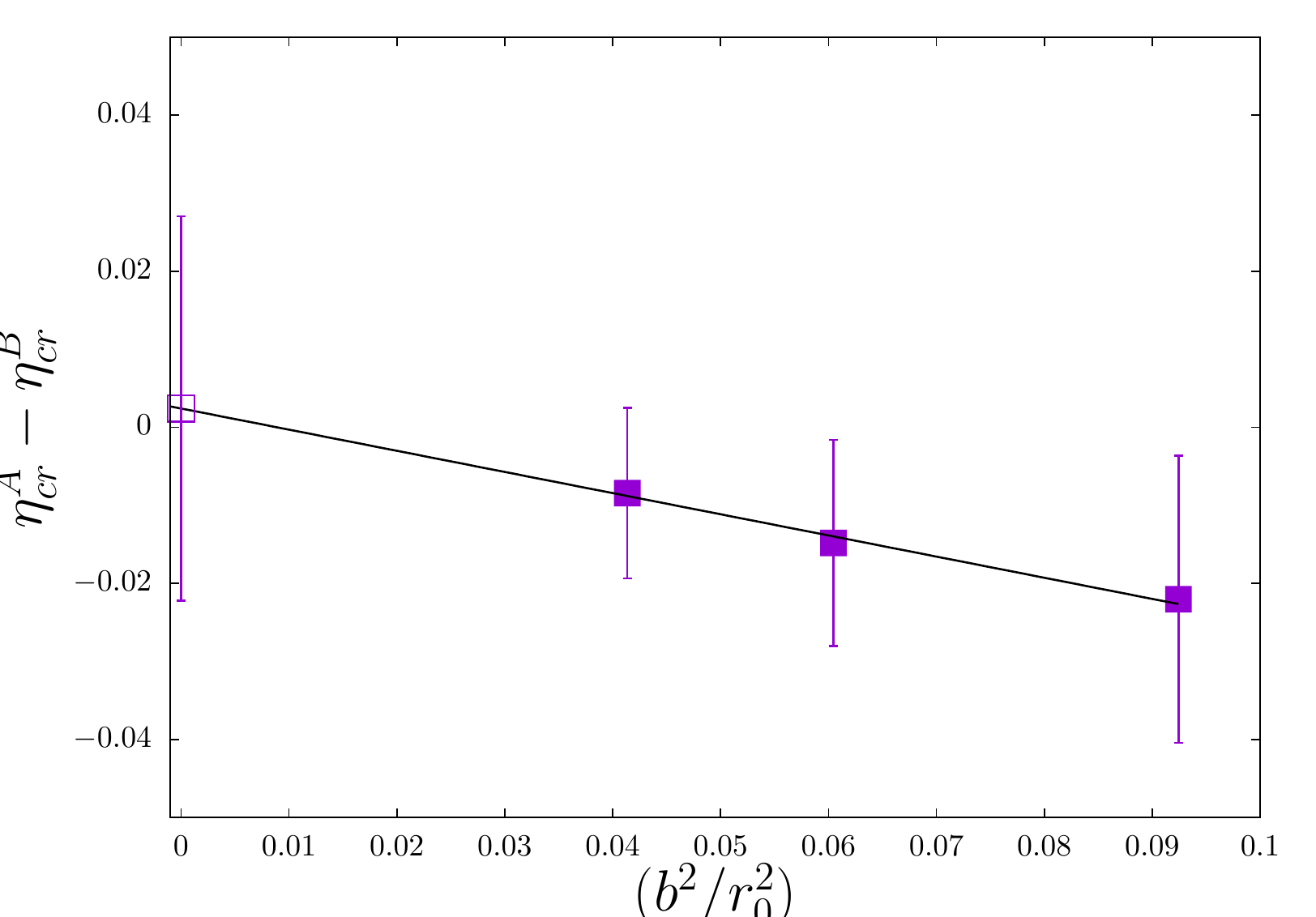}
\caption{}
\label{fig:eta_cr_AB_vs_b}
\end{subfigure}
\caption{(a) typical plot of $r_{AWI}$ against $x_0/b$, here for the case of $\eta=-1.1374$, $b\mu=0.0224$ and $\beta=5.85$. The two time regions where $r_{AWI}$ shows a plateau are $A: x_0 \sim[0.9,1.8]$~fm and $B: x_0 \sim[2.7,3.3]$~fm.
(b) $r_{AWI}$ as a function of $\eta$ at $\beta=5.75$, 5.85 and 5.95. The values of $\eta_{cr}$ at which $r_{AWI}=0$ are denoted by red squares and are $\eta= -1.271(10)
,$ $ -1.207(8)
$ and $ -1.145(6)
$ respectively.
(c) The difference $\eta_{cr}^A - \eta_{cr}^B$ vs.\ $b^2/r_0^2$, where 
$\eta_{cr}^{A(B)}$ is estimated by using the plateau value of $r_{AWI}$ in the 
time region A(B). The value of $\eta_{cr}^A - \eta_{cr}^B$ extrapolated to the 
continuum limit is well compatible with zero.}
\end{figure}

To determine the critical value of  the bare coupling $\eta$ at which the $\tilde \chi$-symmetry is restored  we
look, in the Wigner phase, for the value of $\eta$ at which
\begin{eqnarray}
\hspace{-.8cm}&&r_{AWI}(\eta_{cr},\rho) = \frac{ b^6 \sum_{\bf x,\bf y} \langle P^1(0) 
[ \partial_{0}^{FW} \tilde{A}_0^{1,BW} ](x) \varphi_0(y)\rangle
}{    b^6 \sum_{\bf x, \bf y} \langle P^1(0)
\tilde{D}_P^1 (x) \, \varphi_0(y) \rangle   }\Big{|}_{\eta=\eta_{cr}}= 0 \, ,   \quad \varphi_0\!=\!\frac{1}{2} {\rm Tr}[\Phi] \, .\label{NDRB}
\label{DEF}
\end{eqnarray}
where $\tilde{A}_\mu^{i,BW}(x)=
\frac{1}{2} \overline Q(x-\hat\mu)\gamma_\mu\gamma_5\frac{\tau^i}{2}U_\mu(x-\hat\mu)Q(x) 
+ \frac{1}{2} \overline Q(x)\gamma_\mu\gamma_5\frac{\tau^i}{2}U_\mu^\dagger(x-\hat\mu) Q(x-\hat\mu) $ is the axial current, and
 $P^{i}(x)=\overline Q(x)\gamma_5\frac{\tau^i}{2}Q(x) $,  $D^{i}_P=\overline Q_L\left\{ \Phi,\frac{\tau^i}{2} \right\} Q_R-
\overline Q_R\left\{ \frac{\tau^i}{2},\Phi^\dagger \right\} Q_L$.
The time separation $y_0-x_0$ is kept fixed at $0.6$~fm. The typical profile of $r_{AWI}$ (\ref{NDRB}) is shown in Fig.~\ref{r_AWI_plateau}. It exhibits two plateaux because different intermediate  states dominates the three-point functions in the (\ref{NDRB}) .
 In Fig.~\ref{fig:figRAWI} we plot the extrapolation in $\eta$ of $r_{AWI}$ through $r_{AWI}=0$.
 All
 the values are taken from the time plateau region at $x_0$ smaller than $T /2$
that we call region A.  
We call region B the time plateau region of $r_{AWI}$ corresponds to $x_0$
larger than $T /2$.
The $r_{AWI}$ values extracted
from the time region B are expected to differ from those
coming from the time region A only due to  lattice
artefacts. In fact, we check in Fig.~\ref{fig:eta_cr_AB_vs_b} that the difference
in the values of $\eta_{cr}$ obtained by evaluating $r_{AWI}$ from
the two plateaus of regions A and B vanishes as  $\mbox{O}(b^2 )$
 in the continuum limit.

\subsection{NG phase:}
to check whether the conjectured mass generation actually occurs we compute in the NG phase on a lattice of 
size $L=T/2\sim2.4$~fm  
the effective quark PCAC mass as obtained from the axial Ward identity 
\begin{equation}
\hspace{-.4cm}
m_{AWI}^R\equiv\frac{Z_{\tilde A}}{Z_{P}} \, m_{AWI}
= \frac{Z_{\tilde A} \sum_{\bf x}\partial_0\langle \tilde A_{0}^i(x) P^i(0)\rangle}{2Z_{\tilde P} \sum_{\bf x} \langle P^i(x)P^i(0)\rangle} \Big{|}_{\eta_{cr}}\!, \;\; P^i\!\!=\!\bar Q\gamma_5 \frac{\tau^i}{2}Q\, , \!\!
\label{MAWIR}
\end{equation} 
For practical reasons $Z_P^{-1}$ is evaluated in a hadronic scheme defined   
in the Wigner phase by taking $Z_P^{-1} =r_0^2 G_{PS}^W=r_0^2\langle 0| P^1 |{\rm PS\;meson} \rangle^W$, while  we can trade  $Z_{\tilde{A}}$ with $Z_{\tilde{V}}$ due to  $\chi_L\times\chi_R$ symmetry \cite{Capitani:2017trq}.
An another check that the effective PCAC mass
$\frac{Z_{\tilde A}}{Z_P} m_{AWI}$ is non-zero thus indicates
the presence in the effective lagrangian of a fermion mass term, 
is the fact of the mass of the lowest lying 
pseudoscalar meson state ( computed from the $\langle P^1P^1\rangle$ correlator) is not vanishing.
The propagation of the uncertainty of $\eta_{cr}$ on all the observables gives the dominant contribution to the total error.

\begin{figure}[htbp]
\begin{subfigure}{.5\textwidth}
\includegraphics[scale=0.55,angle=0]{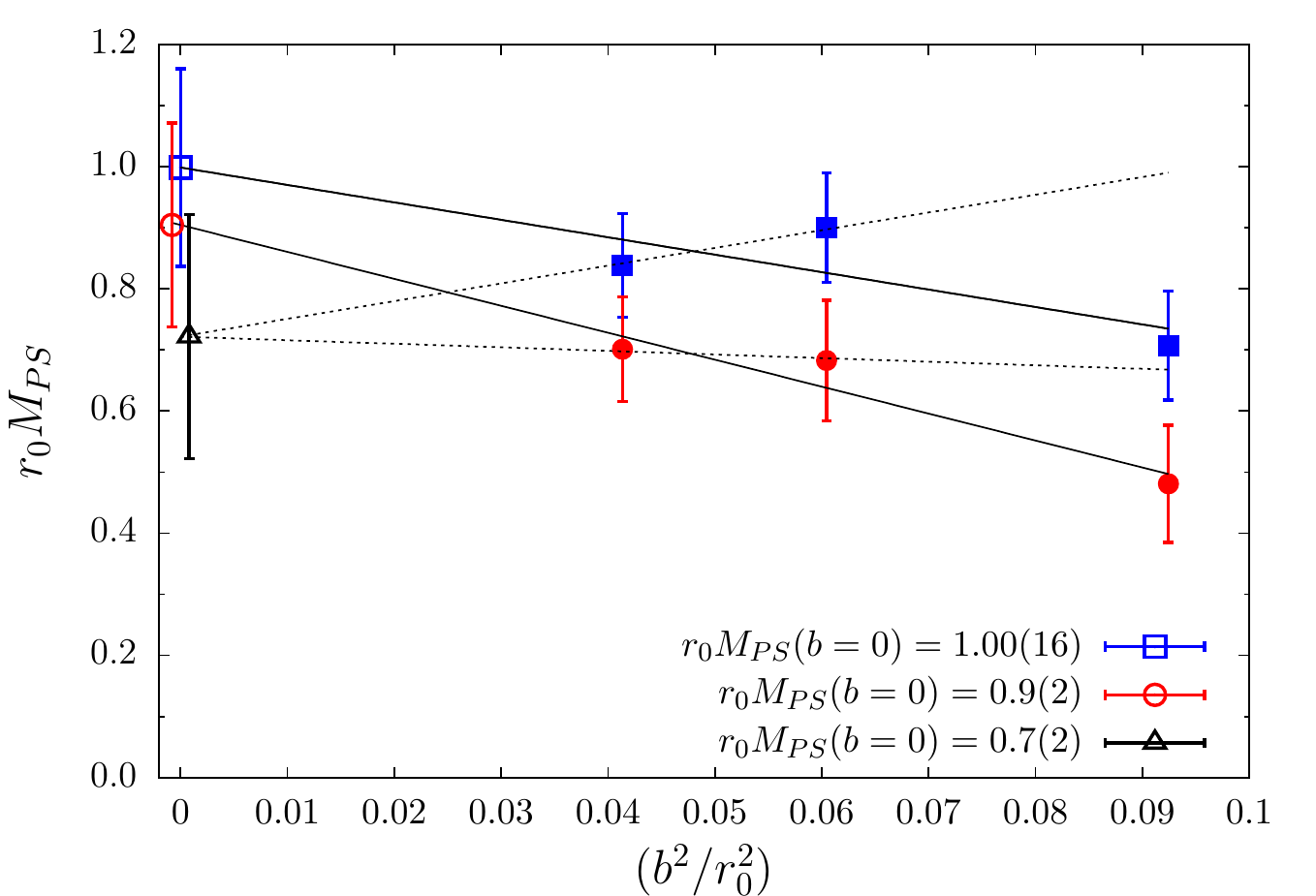} 
\caption{}
\label{fig:M_PS_3ancomp}
\end{subfigure}
\begin{subfigure}{.5\textwidth}
\centerline{\hspace{0.0cm} \includegraphics[scale=0.55,angle=0]{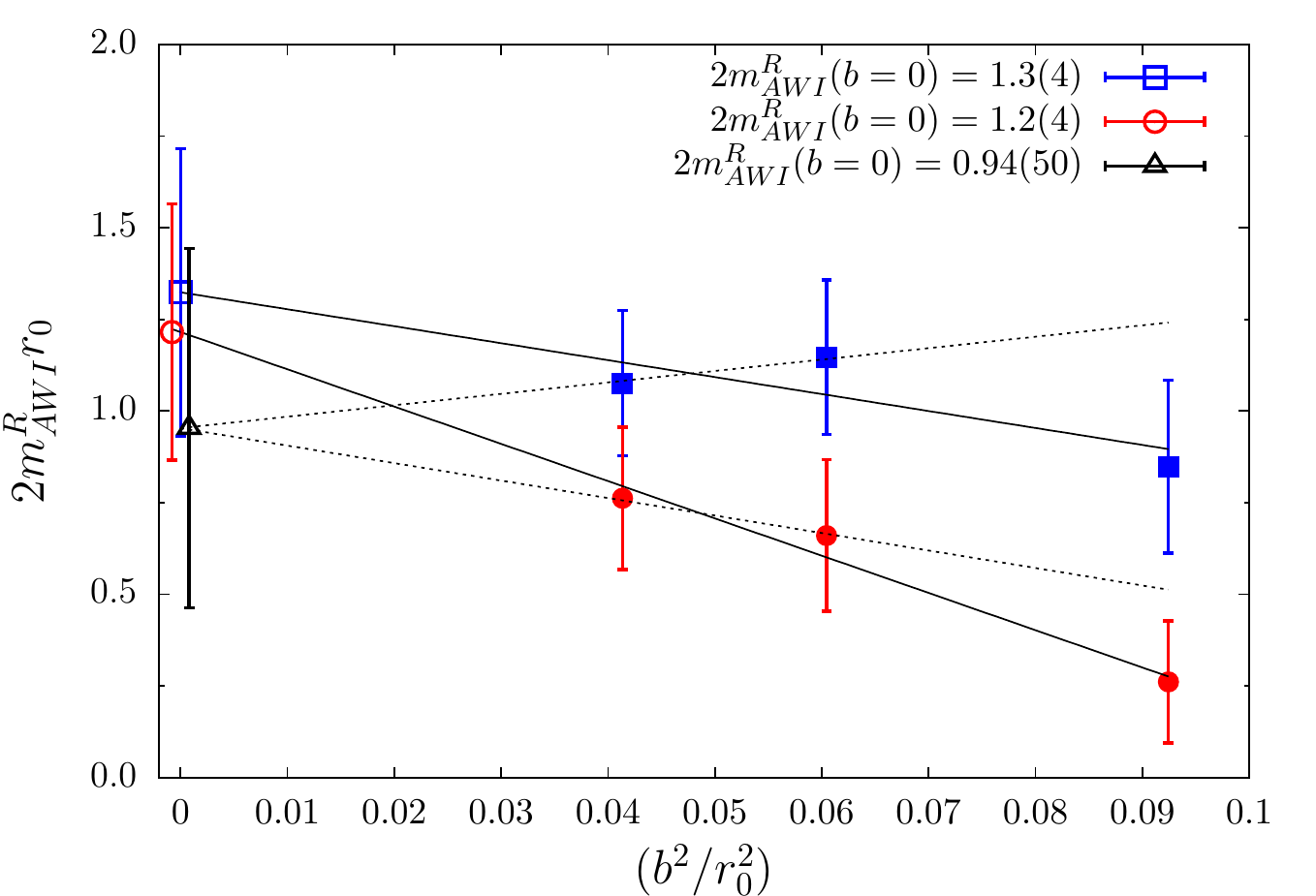} }
\caption{
}
\label{fig:m_AWI_3ancomp}
\end{subfigure}
\caption{(a) Continuum limit extrapolations of $M_{PS}r_0$ data as obtained by working 
at $\eta^A_{cr}$ (blue squares) or at $\eta^B_{cr}$ (red circle) and using data at three lattice spacings as well
as combining data on the two finer lattice at both $\eta^A_{cr}$ and $\eta^B_{cr}$
under the constraint of a common continuum limit. A conservative choice of the time plateau
for $M_{PS}$ and the fit ansatz $M_{PS}^2 =Y_0 + Y_1 \eta + Y_2 \mu +Y_3\eta^2+Y_4\mu^2+Y_5\eta\mu$ were adopted. (b) Data and continuum limit extrapolations of $2m_{AWI}^R r_0 \equiv 2r_0 m_{AWI} 
Z_{\tilde V} Z_P^{-1}$ for the same parameters choice as in (a). The
fit ansatz $m_{AWI}=Y_0+Y_1\eta+Y_2\mu+Y_3\mu^2$ was adopted.
}
\end{figure}

Both the PCAC mass and $M_{PS}$ have been computed for several values of $\mu$ and $\eta$. These values have been extrapolated to $\mu\to0$ and interpolated to $\eta=\eta_{cr}$. In order to have a control over the systematics in our simulation we 
have repeated our analysis modifying:
\begin{enumerate}
\item the plateau choice for $r_{AWI}$, which was taken in the time region either A or B of Fig.~\ref{r_AWI_plateau};
\item the fit function for $r_{AWI}$ in the Wigner phase, which was chosen to be either of the form $K_0+K_1\eta+K_2\mu$ or extended by adding the non-linear term $K_4 \mu^2$ (data are very well linear in $\eta$). In addition the average over the results from the two fits above was considered;
\item the choice of the time plateau for $M_{PS}$ and $m_{AWI}$, taken to be either $[1.4,2.2]$~fm or $[1.1,2.4]$~fm;
 \item the fit functions for $m_{AWI}$ and $M_{PS}^2$ in the NG phase, which 
 were both either truncated to $Y_0+Y_1\eta+Y_2\mu$ 
 or extended by adding in turn terms non-linear in $\eta$ and/or $\mu$,
 such as $Y_4\mu^2$, $Y_3\eta^2 + Y_4\mu^2$ or $Y_3\eta^2 + Y_4\mu^2 + Y_5\eta\mu$.
\end{enumerate}
For each combination of the above choices 
$r_0M_{PS}$ and $2 r_0 m_{AWI} Z_{\tilde V} Z_P^{-1}$ were extrapolated linearly in $b^2/r_0^2$ to the continuum limit.
To estimate the impact of possible O($b^4$) and higher order lattice artifacts, we have 
also considered a continuum extrapolation excluding the coarsest lattice spacing, but 
imposing that the two choices of taking either $\eta^A_{cr}$ or $\eta^B_{cr}$ yield 
the same continuum value.
For a typical choice of the time plateau and the fit function 
in $\eta$ and $\mu$ for $M_{PS}r_0$, in Fig.~\ref{fig:M_PS_3ancomp} we show the 
continuum limit extrapolations obtained by working at $\eta^A_{cr}$ or at $\eta^B_{cr}$
and using data at three lattice spacings, as well as the combined continuum extrapolation
that results from using data on the two finer lattices at both $\eta^A_{cr}$ and $\eta^B_{cr}$.
The analogous comparison for the renormalized ratio $2r_0 m_{AWI} Z_{\tilde V} Z_P^{-1}$ 
is presented in Fig.~\ref{fig:m_AWI_3ancomp} .

The combination of the analysis variants discussed above produced 
72 different analyses, none of which
gave a result for $m_{AWI}$ or $M_{PS}$ that is compatible with zero 
(i.e.\ with the no mechanism scenario). The median over all the analyses, 
excluding only  a few cases where $M_{PS}^2$  yields 
a $\chi^2$ value per degree of freedom larger than 2, gives
\begin{equation}
r_0M_{PS} =\, 0.93\!\pm \!{0.09}\!\pm\!{0.10}\to 0.93(14)\,, \quad \quad
 r_0m_{AWI}^R=  \, 1.20\!\pm \!{0.39}\!\pm\!{0.19}\to 1.20(43)
\end{equation}
where the first error is statistical and the second systematic while the error in bracket is their sum in quadrature. Error  
were estimated by assigning equal weight to all analyses, following the method of 
Ref.~\cite{ETMC_analysis}. 

Further strong evidence for a non-zero fermion mass in the EL comes from 
the following ``{\it reductio ad absurdum}'' argument. Let us assume that no 
fermion mass term is generated.
If this were the case, at $\eta=\eta_{cr}$ 
one should have $M_{PS} = {\mbox{O}}(b^2)$, and hence $M_{PS}^2$
would approach a vanishing continuum limit with only O($b^4$) artifacts.
\begin{figure}[htbp]
{\centerline{\includegraphics[scale=0.6,angle=0]{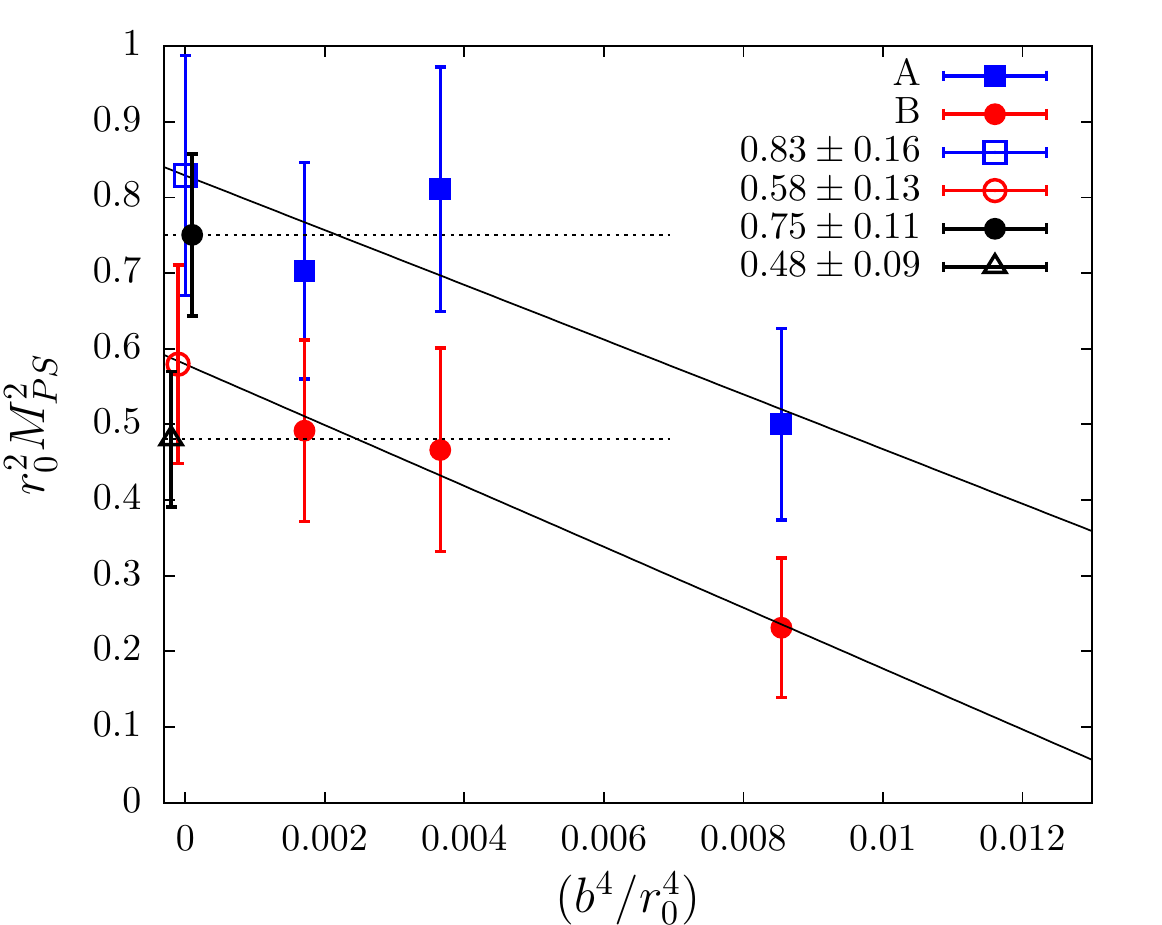}}}
\caption{\small{$r_0^2 M_{PS}^2$ against $(b/r_0)^4$ in the NG phase: 
full (dotted) straight lines show the continuum extrapolation linear in $b^4$ using 
data at all three (only the two finest) lattice spacings. Blue square are the values computed at $\eta_{cr}^A$ while the red circle are the value at $\eta_{cr}^B$ }}
\label{fig:figBQ}
\end{figure}
In Fig.~\ref{fig:figBQ} we plot $r_0^2 M_{PS}^2$ as a function of $(b/r_0)^4$ working at $\eta_{cr}^A$ and $\eta_{cr}^B$.
On a $(b/r_0)^4$ scale the lattice results for $r_0^2 M_{PS}^2$ 
lie very close to the continuum limit, which appears to be non-zero 
by more than five standard deviations. Thus the hypothesis of ``no mechanism'' 
is not supported by lattice data.

\subsection{Numerical investigation at larger values of $\rho$: }
\begin{figure}[hpt]
\begin{subfigure}{.5\textwidth}
\includegraphics[scale=0.5]{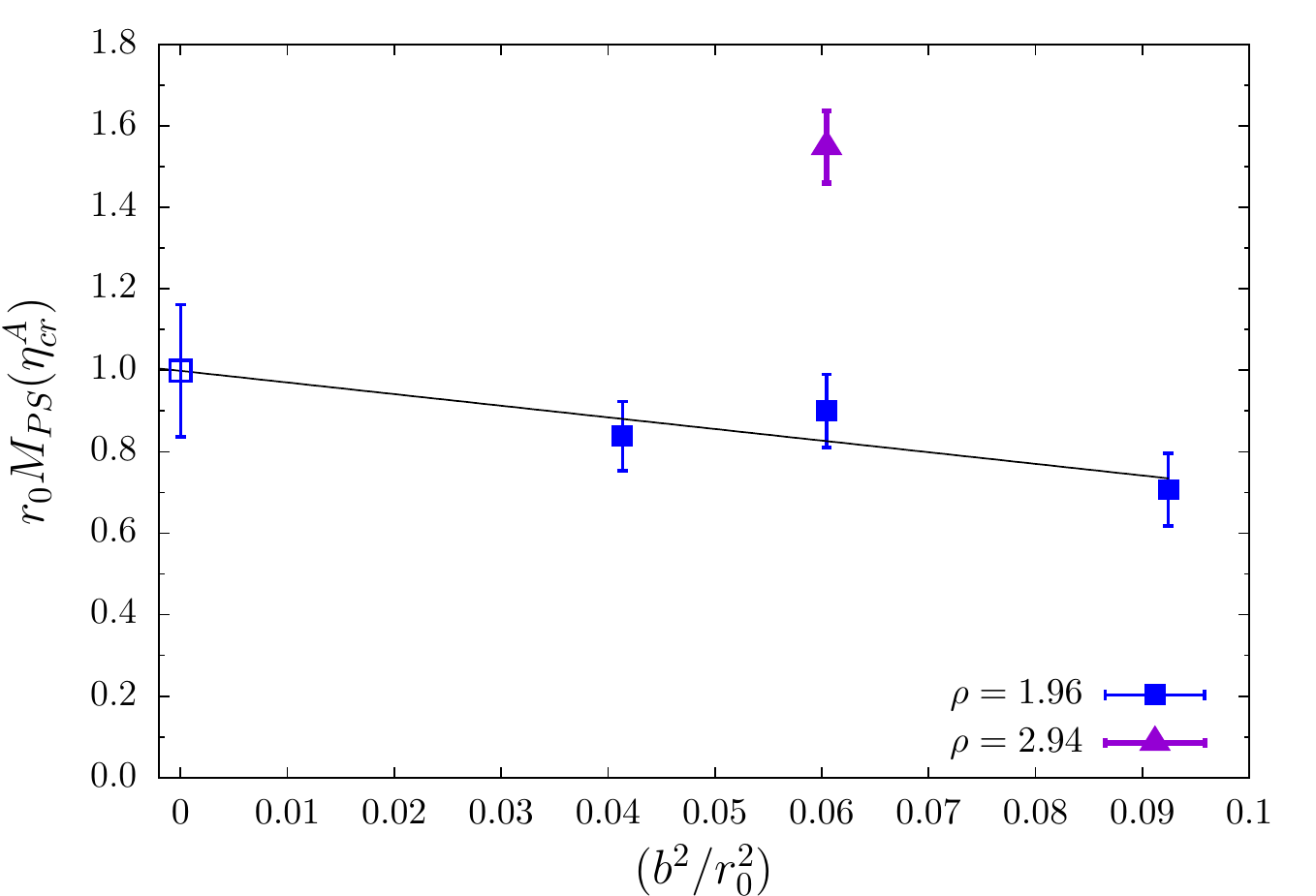}
\caption{}
\label{fig:M_PS_big_rho}
\end{subfigure}
\begin{subfigure}{.5\textwidth}
\includegraphics[scale=0.5]{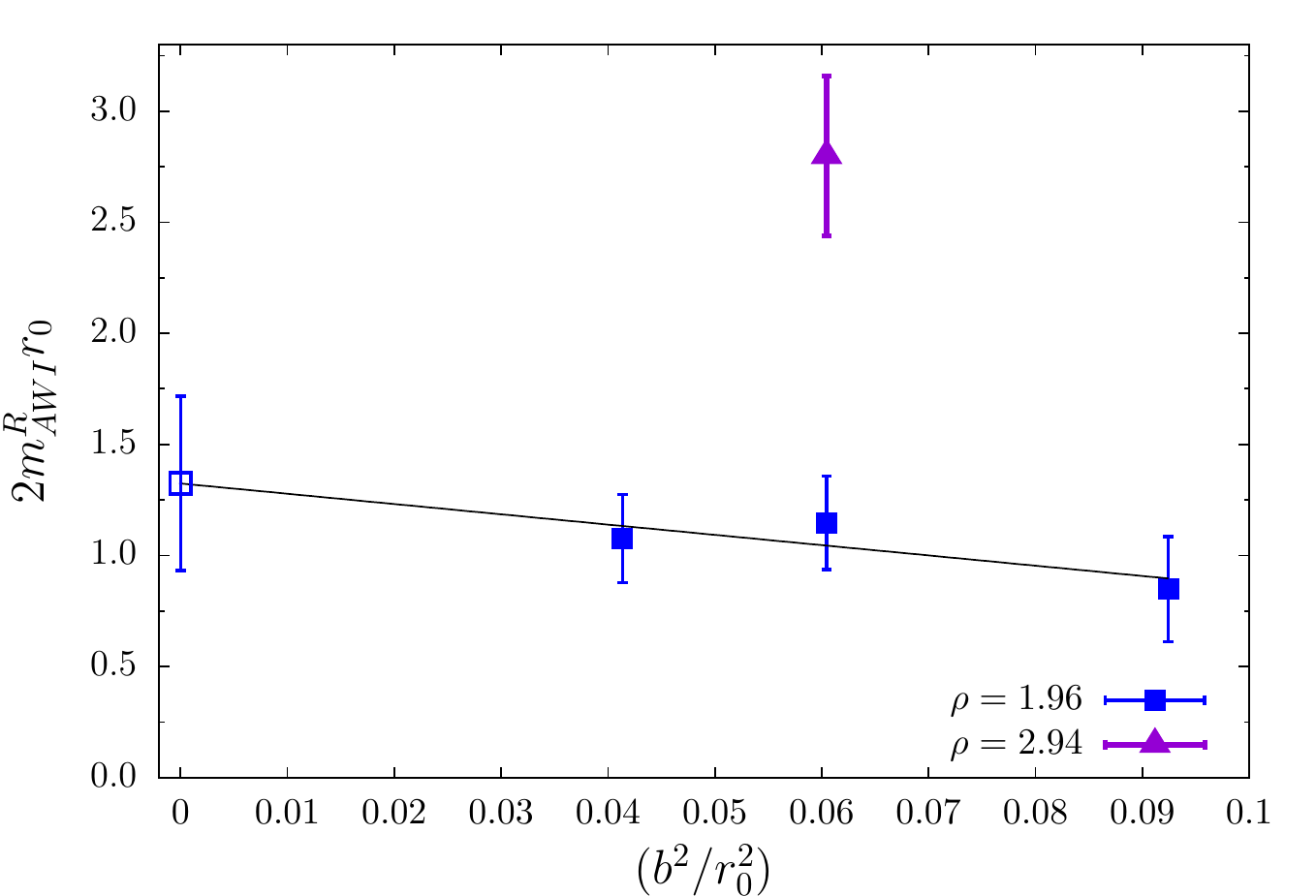}
\caption{}\label{fig:m_AWI_big_rho}
\end{subfigure}
\caption{(a)
Data and continuum limit extrapolations  of $M_{PS}r_0$ as obtained in the NG phase 
at $\eta^A_{cr}$. A conservative choice of the time plateau
for $M_{PS}$ and the fit ansatz $M_{PS}^2 =Y_0 + Y_1 \eta + Y_2 \mu +Y_3\eta^2+Y_4\mu^2+Y_5\eta\mu$ were adopted. Open blue squares refer to $\rho=1.96$ and the violet triangle to $\rho=2.94$.
(b) Data and continuum limit extrapolations of $2m_{AWI}^R r_0 \equiv 2r_0 m_{AWI} 
Z_{\tilde V} Z_P^{-1}$  obtained 
at $\eta^A_{cr}$ in the NG phase. The
fit ansatz $m_{AWI}=Y_0+Y_1\eta+Y_2\mu+Y_3\mu^2$ was adopted. Open blue squares refer to $\rho=1.96$ and the dark violet triangle to $\rho=2.94$.}
\end{figure}

Having found non vanishing values of $m_{AWI}$ and $M_{PS}$
in the continuum limit at $\eta_{cr}$ in the NG phase, as 
conjectured in Ref.~\cite{Frezzotti:2014wja}, we check the behaviour of this masses as a function of $\rho$.
In Ref.~\cite{Frezzotti:2014wja} was conjectured that 
 $m_{AWI}$ and $M_{PS}^2$ are increasing functions of $\rho$
and for small $\rho$ they should increase  at least as O($\rho^2$).
In order to check the $\rho$ dependence of $m_{AWI}$ and $M_{PS}^2$ we have thus  studied
the case of a 1.5 times larger $\rho$ value, namely $\rho = 2.94$, 
for the intermediate lattice spacing ($\beta=5.85$, $a \sim 0.123$~fm) while keeping the scalar sector parameters
fixed to the values given in \S\ref{Numerical investigation}.
In the Wigner phase,
we found that changing $\rho$ from $1.96$ to $2.94$ leads to a substantial change
of $\eta_{cr}$, from $-1.207(8)$ to $-1.838(13)$.
In the NG phase we observe as expected a significant increase of $2r_0m_{AWI} Z_{\tilde A} Z_P^{-1}$ from $1.5(0.21)$ to $2.80(0.36)$ Fig.~\ref{fig:m_AWI_big_rho}, as well as a similar increase of $M_{PS}^2 r_0^2$ Fig.~\ref{fig:M_PS_big_rho}. 

\section{Conclusions}
In this contribution we illustrate our numerical study of the toy model (\ref{SULL}). 
Our results  are compatible with the conjecture \cite{Frezzotti:2014wja} of a new 
non-perturbative (NP) mechanism for elementary mass generation.
 Once the bare parameters are chosen so as to ensure maximal restoration
of fermion chiral symmetries and be in the spontaneously
broken phase of the $\chi_L\times \chi_R$ symmetry, a renormalization group invariant fermion
mass of the order of the $\Lambda_s$ parameter  is generated.
This result  establishes the occurrence of
a non-perturbative obstruction ("anomaly") to the recovery of broken
fermionic chiral symmetries, which is
responsible for the dynamically generated fermion mass
term. The latter, being unrelated to the Yukawa operator,
is expected to become independent of the scalar field
expectation value ($v$) in the limit $\Lambda_s /v \to 0$.
Extension of this mechanism to weak interactions and possible phenomenological
implications are discussed in the contribution of R. Frezzotti  \cite{Frezzotti}

\bibliographystyle{JHEP}
\bibliography{lattice2018.bib}

\providecommand{\href}[2]{#2}\begingroup\raggedright\begin{thebibliography}{1}

\bibitem{Frezzotti:2014wja}
R.~Frezzotti and G.~C. Rossi,
  \href{https://doi.org/10.1103/PhysRevD.92.054505}{\emph{Phys. Rev.}
  {\bfseries D92} (2015) 054505}.

\bibitem{Bochicchio:1985xa}
M.~Bochicchio, L.~Maiani, G.~Martinelli, G.~C. Rossi and M.~Testa,
  \href{https://doi.org/10.1016/0550-3213(85)90290-1}{\emph{Nucl. Phys. B}
  {\bfseries 262} (1985) 331}.

\bibitem{Capitani:2017trq}
S.~Capitani \textit{et~al.},
  \href{https://doi.org/10.1051/epjconf/201817508009}{\emph{EPJ Web Conf.}
  {\bfseries 175} (2018) 08009}.

\bibitem{Guagnelli:1998ud}
{\scshape ALPHA} collaboration, M.~Guagnelli, R.~Sommer and H.~Wittig,
  \href{https://doi.org/10.1016/S0550-3213(98)00599-9}{\emph{Nucl. Phys.}
  {\bfseries B535} (1998) 389}.

\bibitem{Necco:2001xg}
S.~Necco and R.~Sommer,
  \href{https://doi.org/10.1016/S0550-3213(01)00582-X}{\emph{Nucl. Phys.}
  {\bfseries B622} (2002) 328}.

\bibitem{ETMC_analysis}
N.~Carrasco \textit{et~al.},
  \href{https://doi.org/https://doi.org/10.1016/j.nuclphysb.2014.07.025}{\emph{Nuclear
  Physics B} {\bfseries 887} (2014) 19 }.

\bibitem{Frezzotti}
R.~Frezzotti and G.~C. Rossi, {\emph{{PoS, 36th International Symposium on
  Lattice Field Theory (Lattice 2018) East Lansing, MI, United States, July
  22-28, 2018}} }.

\end{thebibliography}\endgroup

\end{document}